\documentclass[12pt,superscriptaddress,aps,prd,preprint,showpacs,nofootinbib]{revtex4}
\usepackage[dvips]{graphicx}
\usepackage{amsfonts}
\usepackage{slashed}
\usepackage{xcolor}

\newcommand{\be}{\begin{equation}}
	\newcommand{\en}{\end{equation}}
\newcommand{\ba}{\begin{eqnarray}}
	\newcommand{\ea}{\end{eqnarray}}
\newcommand{\bea}{\begin{eqnarray}}
	\newcommand{\eea}{\end{eqnarray}}

\begin{document}

	\title{On metric-affine bumblebee model coupled to scalar matter}

	\author{J. R. Nascimento}
	\email[]{jroberto@fisica.ufpb.br}
	\affiliation{Departamento de F\'{\i}sica, Universidade Federal da 
		Para\'{\i}ba,\\
		Caixa Postal 5008, 58051-970, Jo\~ao Pessoa, Para\'{\i}ba, Brazil}
	
	\author{Gonzalo J. Olmo}
	\email[]{gonzalo.olmo@uv.es}
	\affiliation{Departamento de F\'{i}sica Te\`{o}rica and IFIC,
		Centro Mixto Universitat de Val\`{e}ncia--CSIC, Universitat
		de Val\`{e}ncia, Burjassot-46100, Val\`{e}ncia, Spain}
	\affiliation{Universidade Federal do Cear\'a (UFC), Departamento de F\'isica,\\ Campus do Pici, Fortaleza - CE, C.P. 6030, 60455-760 - Brazil.}

	\author{A. Yu. Petrov}
	\email[]{petrov@fisica.ufpb.br}
	\affiliation{Departamento de F\'{\i}sica, Universidade Federal da Para\'{\i}ba,\\
		Caixa Postal 5008, 58051-970, Jo\~ao Pessoa, Para\'{\i}ba, Brazil}
	
	\author{P. J. Porf\'{\i}rio}
	\email[]{pporfirio@fisica.ufpb.br}
	\affiliation{Departamento de F\'{\i}sica, Universidade Federal da 
		Para\'{\i}ba,\\
		Caixa Postal 5008, 58051-970, Jo\~ao Pessoa, Para\'{\i}ba, Brazil}

	
	\begin{abstract}
		We consider the coupling of the metric-affine bumblebee gravity model to scalar matter and calculate the lower-order contributions to two-point functions of bumblebee and scalar fields in the weak gravity approximation. We also obtain the one-loop effective potentials for both scalar and vector fields.
	\end{abstract}
	
	\pacs{11.30.Cp}
	
	\maketitle

	\section{Introduction}
	\label{sec:intro}
	
	Lorentz invariance is generally regarded as the most important symmetry in our current description of the laws of nature. Nonetheless, and despite being a fundamental symmetry,  it is expected to break down at some high-energy scale (typically assumed of Planckian order). This idea was first addressed in the seminal paper \cite{Kos1}, where it was proposed that the Lorentz violating terms may emerge as vacuum expectation values (VEV) of tensor-valued fields in the low-energy limits of effective string theories \cite{Kos1,Kos2,Kos3,Kos4,Kos5}. Lorentz symmetry breaking also arises in other contexts, such as associated to quantum gravity effects \cite{QG, Cal, Boj, Al, Ame} or in extensions of the standard model of particle physics (SM) \cite{Coll1, Coll2, Kos6}.
	
	As is well known, effective field theories (EFT) are powerful tools to probe in the low-energy regime the effects of new physics at the Planck scale \cite{Georgi, Weinberg}. The new effects should manifest themselves as small deviations from known physics caused by  Planck-scale suppressed terms in EFTs as a result of some mechanism taking place in more fundamental theories. In the context of LSB, the most appealing example is the Standard-Model Extension (SME) \cite{Coll1, Coll2}, which is set up as a general effective field framework that incorporates all Lorentz-violating (LV) coefficients (and corresponding operators). Moreover, this framework has been extended to the pure gravitational  sector in \cite{Kos6}. It is worth stressing, in this sense, that the consistent way to implement Lorentz symmetry breaking (LSB) in gravity theories must be different from non-gravity ones. In fact, in flat spaces, requirements for vectors (or, generically, tensors)  to be constant are implemented by simple conditions, such as $\partial_{\mu}k_{\nu}=0$, where $k_{\mu}$ is an arbitrary non-trivial vector VEV. On the other hand, their natural covariant generalizations, like $\nabla_{\mu}k_{\nu}=0$, do not yield the suitable manner to define LSB in curved spaces because such a condition leads to restrictive constraints on the allowed spacetime geometries (no-go constraints) \cite{KosLi}. The way out to solve this problem is to allow the Lorentz violation to be driven by dynamical vacuum expectation values (VEVs) which, in turn, should satisfy their own equations of motion. A particular example of this is the bumblebee gravity models \cite{Casana, Bertolami, Santos1, Santos2, Santos3, Maluf1, Maluf2, Kumar}.
	
	The pure-gravity sector of the SME has been built in a Riemann-Cartan background, which allows for the presence of dynamical torsion \cite{Kos6}. This contrasts with the fact that in most of the works in the literature dealing with modified theories of gravity that incorporate LSB in (pseudo)-Riemannian spacetimes, i.e., within the usual metric framework of gravity -- the metric is the only dynamical geometric field. In this scenario, one can cite, for example, the Einstein-aether model \cite{Jacobson}, the bumblebee gravity \cite{Casana, Bertolami, Santos1, Santos2, Santos3, Maluf1, Maluf2, Kumar} and Chern-Simons modified gravity \cite{Jackiw}. Therefore, considering geometric frameworks more general than the metric approach can provide new insights into the phenomenological aspects of LSB theories in the presence of gravity. Along this line, modified theories of gravity with LSB have been deemed in different non-Riemaniann geometric scenarios, such as in Finsler geometry \cite{Foster1, Schreck1, Schreck2, Edwards, Mdo} and in the metric-affine approach\footnote{In this approach, metric and connection are assumed to be independent geometrical entities a priori.} \cite{PJ1, PJ2, PJ3, PJ4, Boudet1, Boudet2}.

	From the phenomenological perspective, considering particular modified theories of gravity that lead to LSB seems to be a promising way to explore the signals of new physics by confronting the theoretical predictions with experimental/observational data \cite{datatables}. In this respect, the bumblebee gravity, introduced in \cite{KosLi}, represents a realistic model that incorporates the effects of LSB through a dynamical mechanism of spontaneous symmetry breaking. In this model, the spontaneous LSB is driven by the potential of
	a vector field, $B_{\mu}$, which acquires a non-trivial  VEV  \cite{KosLi}. As discussed above, although most studies of LSB in modified theories of gravity have been performed in the traditional metric approach,  a metric-affine version of the bumblebee model has been proposed and explored in some detail \cite{PJ1, PJ2, PJ3}, obtaining promising results both at the classical \cite{PJ1} and quantum levels \cite{PJ2,PJ3}. In this sense, while in \cite{PJ1} the coupling of spinor and scalar matter fields was considered, only perturbative effects of spinor fields were further studied at the quantum level in  \cite{PJ2,PJ3}. We note that within the quantum field theory, perturbative studies of any new field theory model are carried out through its coupling to some standard fields, namely, scalar, spinor and gauge ones, for example, in the QED the gauge field is coupled to scalar or spinor ones. While the coupling of the bumblebee gravity to the spinor field has been already studied perturbatively in \cite{PJ1, PJ2, PJ3}, its interaction with scalar field was not considered yet. So, its investigation is a natural problem. The quantum dynamics of scalar fields coupled to metric-affine bumblebee gravity thus remains an open question, and that is the point we pursue in this work. 
	
	This paper is organized as follows. In section \ref{review}, we provide a brief review of the metric–affine bumblebee gravity. In section \ref{weak}, we investigate the weak field limit of the metric–affine bumblebee gravity minimally coupled with a scalar field, and then we find an explicit expression for the scalar field propagator. Additionally, we also find the contributions to the two-point functions at the one-loop level for the effective action. In section \ref{potential}, we computed the effective potential for both the vector and scalar fields. Finally, in section \ref{con}, we conclude the paper by providing a summary and some conclusions.
	
	\section{The metric-affine bumblebee gravity}\label{review}

	In this section, we review the main features of the metric-affine bumblebee gravity. The action of the model is given by \cite{PJ1, PJ2, PJ3}
	\begin{eqnarray}
		\nonumber \mathcal{S}&=&\int d^4 x\,\sqrt{-g}\Big[\frac{1}{2\kappa^2}\Big(R(\Gamma)+\xi B^{\alpha} B^{\beta} R_{\alpha\beta}(\Gamma)\Big)-\frac{1}{4}B^{\mu\nu}B_{\mu\nu}-V(B^{\mu}B_{\mu}\pm b^2)\Big] +\\
		&+&\int d^4 x\,\sqrt{-g}\mathcal{L}_{m}(g_{\mu\nu},\Gamma^{\mu}_{\alpha\beta},\Psi_i),
		\label{bumblebee}
	\end{eqnarray}
	where $R_{\mu\nu}(\Gamma)=R^{\alpha}_{\,\,\,\mu\alpha\nu}(\Gamma)$ is the Ricci tensor of the independent connection $\Gamma^\alpha_{\mu\nu}$, $R(\Gamma)=g^{\mu\nu}R_{\mu\nu}$ is the Ricci scalar, $\mathcal{L}_{m}(g_{\mu\nu},\Gamma^{\mu}_{\alpha\beta},\Psi_i)$ stands for the Lagrangian of the contributions stemming from the matter sources, which are represented by the set of generic fields $\Psi_{i}$ and $\kappa^2 =8\pi G$. The coupling constant $\xi$ controls the non-minimal interaction between the bumblebee field, $B_{\mu}$, and the curvature, as shown in the above action. In addition, the field strength is defined by $B_{\mu\nu}=\left(dB\right)_{\mu\nu}$. The bumblebee potential, $V$, possesses a non-trivial minimum, we say at $B_{\mu}=b_{\mu}$, where $b_{\mu}$ is the bumblebee vacuum expectation value (VEV), and $b^2=g^{\mu\nu}b_{\mu}b_{\nu}>0$. At the minimum, the (local) Lorentz symmetry is spontaneously broken. It is noteworthy that the aforementioned action is treated within the metric-affine approach in which the metric and connection are taken to be independent geometrical quantities. 
	
	For our purposes in this work, we shall consider a scalar field as the only matter source minimally coupled to gravity. In this case, the scalar field just couples to the metric, while their coupling with the connection is completely factored out, thus $\mathcal{L}_{m}(g_{\mu\nu},\Gamma^{\mu}_{\alpha\beta},\Psi_i)=\mathcal{L}_{m}(g_{\mu\nu},\Psi_i)$. Putting this together with the fact that only the symmetric part of the Ricci tensor plays a role in the action (\ref{bumblebee}), one concludes that the action of this model is invariant under projective transformations; as a result, ghost-like propagating degrees of freedom are prevented in the gravitational sector of the model \cite{Adria}. 
	
	\subsection{Field equations}
	
	We obtain the field equations by varying Eq.(\ref{bumblebee}) with respect to the metric, connection and bumblebee field, respectively,
	\begin{eqnarray}
		\kappa^2  T_{\mu\nu}&=&R_{(\mu\nu)}(\Gamma)-\frac{1}{2}g_{\mu\nu}\bigg(R(\Gamma)+\xi B^{\alpha}B^{\beta}R_{\alpha\beta}(\Gamma)\bigg)+2\xi\bigg(B_{(\mu}R_{\nu)\beta}(\Gamma)\bigg)B^{\beta};\label{Riccieq}\\
		0&=&\nabla_{\lambda}^{(\Gamma)}\left[\sqrt{-g}g^{\mu\beta}\left(\delta^{\nu}_{\,\,\beta}+\xi B^{\nu}B_{\beta}\right)\right];\label{connectioneq}\\
		\nabla_{\mu}^{(g)}B^{\mu}_{\ \nu}&=&-\frac{\xi}{\kappa^2}B^{\beta}R_{\beta\nu}(\Gamma)+2 V^{\prime}B_{\nu},\label{bumblebeeeq}
	\end{eqnarray}
	where we have defined $T_{\mu\nu}=T_{\mu\nu}^{(m)}+T_{\mu\nu}^{(B)}$, with
	\begin{eqnarray}
		T_{\mu\nu}^{(m)}=-\frac{2}{\sqrt{-g}}\frac{\delta\left(\sqrt{-g}\mathcal{L}_M\right)}{\delta g^{\mu\nu}},
	\end{eqnarray}
	is the stress-energy tensor of the matter sources and
	\begin{equation}
		T_{\mu\nu}^{(B)}=B_{\mu\sigma}B_{\nu}^{\,\,\sigma}-\frac{1}{4}g_{\mu\nu}B^{\alpha}_{\,\,\sigma}B^{\sigma}_{\,\,\alpha}-Vg_{\mu\nu}+2V^{\prime}B_{\mu}B_{\nu},
	\end{equation}
	is the stress-energy tensor of the bumblebee field and $V^{\prime}$ stands for derivative with respect to its argument. It should be emphasized that we have got rid of the torsional terms by using the gauge (projective) freedom in the connection equation (\ref{connectioneq}). The solution to this equation is simply given by the Christoffel symbols of the auxiliary metric ($h_{\mu\nu}$), namely,
	\begin{equation}
		\Gamma^{\alpha}_{\mu\nu}=\frac{1}{2}h^{\alpha\beta}\left(-\partial_{\beta}h_{\mu\nu}+\partial_{\mu}h_{\nu\beta}+\partial_{\nu}h_{\mu\beta}\right),
		\label{conn}
	\end{equation}
	where
	\begin{eqnarray}
		h^{\mu\nu}&=&\frac{1}{\sqrt{1+\xi X}}\left(g^{\mu\nu}+\xi B^{\mu}B^{\nu}\right);\\
		h_{\mu\nu}&=&\sqrt{1+\xi X}\left(g_{\mu\nu}-\frac{\xi}{1+\xi X}B_{\mu}B_{\nu}\right),  
	\end{eqnarray}
	with $X\equiv g^{\mu\nu}B_{\mu}B_{\nu}$. For a detailed step-by-step derivation of these results, see \cite{our}. 
	
	The action (\ref{bumblebee}) admits an Einstein frame representation, it can be seen by integrating out the connection by using Eq.(\ref{conn}), doing so, we get
	\begin{equation}
		\tilde{S}=\frac{1}{2\kappa^2}\int d^{4}x\sqrt{-h}\,R(h)+\int d^{4}x\sqrt{-h}\,\mathcal{L}_{m}(h_{\mu\nu}, B_{\mu}, \Psi_{i})+ \tilde{S}_{B},
		\label{Eframe}
	\end{equation}
	where $\tilde{S}_{B}$ represents the vectorial sector of the model in the Einstein frame. Physically speaking, the gravitational sector of Eq.(\ref{Eframe}) is simply the Einstein-Hilbert action for the auxiliary metric $h_{\mu\nu}$ plus a modified matter action $\mathcal{L}_{m}(h_{\mu\nu}, B_{\mu}, \Psi_{i})$, which carries non-linear interactions among $h_{\mu\nu}$, $B_{\mu}$ and the matter fields. In practical terms, the non-linear interactions between the bumblebee field and Ricci tensor in (\ref{bumblebee}) have been shifted to the matter sector, as viewed from the perspective of the Einstein frame (\ref{Eframe}). Finally, as pointed out in \cite{ PJ2}, the non-metricity tensor $Q_{\mu\alpha\beta}\equiv -\nabla^{(\Gamma)}_{\mu}g_{\alpha\beta}$ is completely sourced by the spontaneous Lorentz symmetry breaking terms, which contrasts with the approach of \cite{Foster} in which LSB is explicitly broken.
	
	Regarding the bumblebee field equation, it can be cast into a Proca-like form, as shown below,
	\begin{equation}
		\nabla^{(g)}_{\mu}B^{\mu\nu}=\mathcal{M}^{\nu}_{\,\mu}B^{\mu},
	\end{equation}
	where $\mathcal{M}^{\nu}_{\,\,\mu}$ is the effective mass-squared tensor, explicitly defined by
	\begin{eqnarray}
		\nonumber\mathcal{M}^{\nu}_{\,\,\mu}&=& \left(\frac{\xi T}{2+3\xi X}+\frac{\xi^{2}B^{\alpha}B^{\beta}T_{\alpha\beta}}{\left(1+\xi X\right)\left(2+3\xi X\right)}+2V^{\prime}\right)\delta^{\nu}_{\,\mu}-
		\frac{\xi}{1+\xi X}T^{\nu}_{\,\mu}.
		\label{massS}
	\end{eqnarray}
	It is evident from the previous equation that the effective massive term ($\mathcal{M}^{\mu}_{\,\,\nu}$) accounts for unconventional new interactions between the bumblebee field and the stress-energy  tensor, apart from the usual purely massive term $V^{\prime}$ and other terms proportional to the stress-energy tensor. The relative sign between the two terms in Eq.(\ref{massS}) allows, in principle, the emergence of instabilities. However, we call attention to the fact that this model should be interpreted within an effective field theory perspective \cite{Georgi}.

	\section{The weak field limit}\label{weak}
	
	Our aim in this section is to investigate particle physics scenarios (weak gravitational field regime), which correspond to taking $h_{\mu\nu}\approx \eta_{\mu\nu}$ in the Einstein frame, as this metric satisfies a set of Einstein-like equations and, therefore, is only sensitive to the integrated distributions of matter and energy, which should be small. The space-time metric $g_{\mu\nu}$ is, in addition, affected by the local energy densities, admitting a weak-field approximation of the form $g_{\mu\nu}\approx \eta_{\mu\nu}+\xi\left(B_{\mu}B_{\nu}-\frac{1}{2}X \eta_{\mu\nu}\right)$ up to first order in $\xi$ \cite{PJ2}. In order to further explore the dynamics of a scalar field non-minimally coupled to gravity in the weak field regime, it is necessary to make explicit the bumblebee potential. Without loss of generality, we choose the traditional Mexican hat-like potential, i.e.,
	\begin{equation}
		V(B^{\mu}B_{\mu}\pm b^2)=\frac{\lambda}{4}(B^{\mu}B_{\mu}\pm b^2)^2,
	\end{equation}
	where $\lambda$ is a positive self-coupling constant. 
	
	In the weak field limit, it has been shown in \cite{PJ2,PJ3} that the effective dynamics is described by the following scalar and bumblebee Lagrangians:
	\begin{eqnarray}
		\nonumber\mathcal{L}_{sc}&=&-\frac{1}{2}\Phi(\Box+m^2)\Phi-\frac{\xi}{2}(B^{\mu}\partial_{\mu}\Phi)^2+\frac{m^2}{4}\xi\Phi^2B^{\mu}B_{\mu}+
		\mathcal{O}(\xi^2),\\
		&=&-\frac{1}{2}\Phi(\Box+m^2)\Phi+\frac{\xi}{2}\Phi\Big[B^{\mu}B^{\nu}\partial_{\mu}\partial_{\nu}+\\
		\nonumber&+& \big(B^\mu (\partial_\nu B^\nu)+B^\nu (\partial_\nu B^\mu)\big)\partial_\mu+\frac{m^2}{2} B^2\Big]\Phi+\mathcal{O}(\xi^2),\\
		\nonumber{\cal L}_{BEF}&=&-\frac{1}{4}B_{\mu\nu}B^{\mu\nu}+\frac{M^2}{2}B^2-\frac{\Lambda}{4}(B^2)^2+\\ \nonumber&+&\frac{\xi}{2}\Big[B^{\mu\nu}B^\alpha{}_{\nu}B_\mu B_\alpha-\frac{1}{4}B_{\mu\nu}B^{\mu\nu}B^2-\frac{3}{4}\Lambda(B^2)^3\Big]+
		\mathcal{O}(\xi^2)
		\label{BumbLagPert}
	\end{eqnarray} 
	respectively, where the bumblebee effective mass is given by
	$M^2= \lambda b^2(\pm 1+\frac{\xi}{4}b^2)$ and $\Lambda\equiv \lambda(1\pm 2\xi b^2)$. Following the same notation of the earlier papers \cite{PJ2,PJ3}, we assume that indices are raised and lowered with the Minkowski metric, so that $B^2=\eta_{\mu\nu}B^\mu B^\nu$ and so on, which is consistent with neglecting $\mathcal{O}(\xi^2)$ terms. 
	
	Here we assume that the spontaneous Lorentz symmetry breaking is generated by the potential, in our case looking like $V=-\frac{M^2}{2}B^2+\frac{\Lambda}{4}(B^2)^2$, since higher-order terms coming from the non-metricity contributions, can be suppressed by different powers of $\xi$. 

	
	Following \cite{Gomes:2007mq}, let us now compute the quantum corrections on top of a stable non-trivial bumblebee VEV characterized by $<B^\mu>=\beta^\mu$, where {$\beta^2=\pm b^2$}. The dynamics of small perturbations of the bumblebee field can then be analyzed by expanding the Lagrangian (\ref{BumbLagPert}) around the vacuum as $B_{\mu}=\beta_{\mu}+\tilde{B}_{\mu}$. This leads to the following Lagrangian for the perturbations
	\begin{eqnarray}
		\nonumber{\cal L}_{BEF}^{\rm pert}&=&-\frac{1}{4}\tilde{\eta}_{\mu\alpha}\tilde{B}^{\mu}_{\,\,\nu}\tilde{B}^{\alpha\nu}-\Lambda(\beta_\mu\tilde{B}^\mu)^2-\frac{\Lambda}{4}(\tilde{B}^2)^2-\Lambda\tilde{B}^2\tilde{B}_{\rho}\beta^{\rho}+\\
		\nonumber&+&\xi\Big[\frac{1}{2}\tilde{B}^{\mu\nu}\tilde{B}^\alpha{}_\nu\tilde{B}_\alpha\tilde{B}_{\mu}-\frac{1}{4}\tilde{B}^{\mu\nu}\tilde{B}_{\mu\nu}(\beta_\rho\tilde{B}^\rho)+\\
		&+&\tilde{B}^{\mu\nu}\tilde{B}^\alpha{}_\nu\tilde{B}_\alpha\beta_{\mu}-\frac{1}{8}\tilde{B}^{\mu\nu}\tilde{B}_{\mu\nu}\tilde{B}^2\Big]+\mathcal{O}(\xi^2,\lambda\xi),
		\label{BumbLagVEV}
	\end{eqnarray}
	where the kinetic term for the perturbations interacts with the background by coupling to the effective metric 
	\begin{eqnarray}
		\tilde{\eta}_{\mu\nu}\equiv\eta_{\mu\nu}\left(1+\frac{\xi\beta^2}{2}\right)-2\xi \beta_{\mu}\beta_{\nu}.
	\end{eqnarray}
	We note that in this case, due to the non-trivial minimum, the Maxwell-like term is rescaled looking like $-\frac{1}{4}\tilde{B}_{\mu\nu}\tilde{B}^{\mu\nu}(1+\frac{\xi\beta^2}{2})$, and an aether-like term arises. 
	
	The free (linearized) equations of motion for the vector field are
	\begin{eqnarray}
		\nonumber 0&=&\partial_{\mu}\tilde{B}^{\mu\nu}\left(1+\frac{\xi\beta^2}{2}\right)-\xi \beta_{\mu}\beta_{\alpha}\partial^{\mu}\tilde{B}^{\alpha\nu}+\xi \beta^{\nu}\beta_{\alpha}\partial_{\mu}\tilde{B}^{\alpha\mu}-
		2\Lambda\beta^{\nu}\beta_{\alpha}\tilde{B}^{\alpha},
	\end{eqnarray}
	or, in terms of the effective metric, they look like
	\begin{equation}
		\tilde{\eta}_{\mu[\nu}\partial^{\nu}\tilde{B}^{\mu}_{\,\,\alpha]}+M_{\alpha\mu}\tilde{B}^{\mu}=0,
	\end{equation}
	where $M_{\alpha\mu}=-2\Lambda\beta_{\alpha}\beta_{\mu}$ is the effective mass-squared tensor.
	Taking the divergence of this equation, which must be zero, we find that, 
	instead of the usual condition $(\partial\cdot B)=0$, we will have an essentially new condition $(\beta\cdot\partial)(\beta\cdot \tilde{B})=0$ (from the formal viewpoint, this condition is explained by the fact that our mass term is also aether-like, but not the usual Proca mass term).
	Under this condition, the free action of the vector field becomes
	\begin{eqnarray}
		{\cal L}&=&-\frac{1}{4}\tilde{B}_{\mu\nu}\tilde{B}^{\mu\nu}\left(1+\frac{\xi\beta^2}{2}\right)-\frac{\xi}{2}\tilde{B}_{\mu}[\Box\beta^{\mu}\beta^{\nu}+
		(\beta\cdot\partial)^2\eta^{\mu\nu}]\tilde{B}_{\nu}-\Lambda(\beta^{\alpha}\tilde{B}_{\alpha})^2.
	\end{eqnarray}
	We assume $\xi$ to be small (recall that it has dimensions of inverse squared mass, so that it can be treated as an inverse square of some large mass scale). So, we can consider corrections of first order in $\xi$ only, which are given by one-vertex graphs and yield contributions to the two-point function. This actually means that we need only quartic vertices. We consider again the graphs given in Fig. 1 which only yield contributions to the two-point function. To calculate those graphs, it remains to write down the background-dependent propagators. For the vector field, the propagator was found in \cite{PJ3} and looks like
	\begin{eqnarray}
		\nonumber<\tilde{B}_{\alpha}(-k)\tilde{B}_{\beta}(k)>&=&i\frac{1}{-k^2(1+\frac{\xi\beta^2}{2})+\xi(\beta\cdot k)^2}\Big[\eta_{\alpha\beta}-
		\Delta^{-1}\Big([-k^2(1-\frac{1}{2}\xi\beta^2)+\xi(\beta\cdot k)^2-\\
		&-&2\Lambda\beta^2](1+\frac{\xi\beta^2}{2})k_{\alpha}k_{\beta}+
		\xi(\beta\cdot k)^2(-2\Lambda+\xi k^2)\beta_{\alpha}\beta_{\beta}-\nonumber\\\nonumber&-&
		(1+\frac{\xi\beta^2}{2})
		(-2\Lambda+\xi k^2)(\beta\cdot k)(\beta_{\alpha}k_{\beta}+\beta_{\beta}k_{\alpha})
		\Big)\Big],
	\end{eqnarray}
	with
	\begin{eqnarray}
		\Delta&=&-(-2\Lambda+\xi k^2)(\beta\cdot k)^2(1+\frac{\xi\beta^2}{2})+\nonumber\\&+&
		\xi(\beta\cdot k)^2[-k^2(1-\frac{1}{2}\xi\beta^2)-2\Lambda\beta^2+\xi(\beta\cdot k)^2].
	\end{eqnarray}
	We see that the exact propagator of the vector field is highly cumbersome.  However, the leading order in the two-point function of the vector field is of zero order in $\xi$, and of first order in the spinor field, since spinor-vector vertices already contain $\xi$.
	Therefore, to consider the lower-order contributions, it is sufficient to take into account  in the propagator only zero order in $\xi$:
	\begin{eqnarray}
		\Delta&=&2\Lambda(\beta\cdot k)^2+O(\xi).
	\end{eqnarray}
	At the same time, after introducing the background $\beta^{\mu}$, the free Lagrangian for the scalar $\Phi$ becomes
	\begin{eqnarray}
		{\cal L}_{sc}&=&-\frac{1}{2}\Phi\left(\Box+m^2-\xi[(\beta\cdot\partial)^2+
		\frac{m^2}{2} \beta^2]\right)\Phi+\mathcal{O}(\xi^2).
	\end{eqnarray}
	
	So, we can write our propagators as follows (cf. \cite{PJ2}):
	\begin{eqnarray}
		\label{propzero}
		\nonumber<\tilde{B}_{\alpha}(-k)\tilde{B}_{\beta}(k)>&=&-i\frac{1}{k^2}\Big[\eta_{\alpha\beta}-
		\frac{1}{2\Lambda(\beta\cdot k)^2}\Big([-k^2-2\Lambda\beta^2]k_{\alpha}k_{\beta}+\\
		\nonumber&+&2\Lambda(\beta\cdot k)(\beta_{\alpha}k_{\beta}+\beta_{\beta}k_{\alpha})
		\Big)\Big]
		+O(\xi);\nonumber\\
		<\Phi(-k)\Phi(k)>&=&\frac{i}{k^2-m^2(1-\frac{\xi\beta^2}{2})-\xi(\beta\cdot k)^2}.
	\end{eqnarray}
	\begin{figure}[h]
		\centering
		\includegraphics[scale=1]{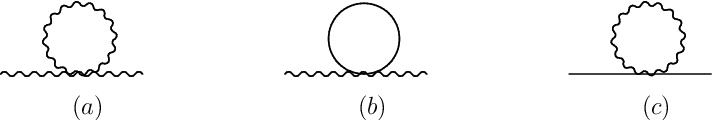}
		\caption{\label{Fig1} The contributions to the two-point functions.}
	\end{figure}
	
	With these propagators, we can find the contributions from graphs (a), (b), (c) depicted in Fig. 1, and also of the graph (i) given by Fig. 2, which we define later. We note that the contributions (a), (b), (c) are present in the trivial vacuum case as well.
	
	The graph (a) is just the same as in the theory where, instead of scalars, spinors are coupled to the bumblebee field. Its calculation was considered in full detail in \cite{PJ2}, where it was shown that this graph leads to vanishing contribution, as it is proportional to identically vanishing tadpole integrals
	\begin{eqnarray}
		\int \frac{d^{d}k}{(2\pi)^d}\frac{1}{k^2}=0, \,\, \int \frac{d^{d}k}{(2\pi)^d}\frac{1}{(\beta\cdot k)^2}=0\label{kk}
	\end{eqnarray}
	and 
	\begin{eqnarray}
		\int \frac{d^{d}k}{(2\pi)^d}\frac{1}{(\beta\cdot k)^n}=0,\,\, \mbox{with}\,\, n=0,1,2...\,\, ,
		\label{kkk}
	\end{eqnarray}
	together with the table of integrals given in \cite{Leib}.  We note that while vanishing of the first integral of (\ref{kk}) is known to be easily proved within the dimensional regularization approach, the vanishing of the above integrals involving the $\beta_{\mu}$ vectors can be demonstrated in any base where the vector $\beta_{\mu}$ is aligned with any of the coordinate axes, say $k_0$, i.e. $\beta\propto(1,0,0,0)$, so, $	\int \frac{d^{d}k}{(2\pi)^d}\frac{1}{(\beta\cdot k)^n}\propto\int\frac{dk_0}{2\pi\beta^2k^n_0}\int d^{d-1}\vec{k}$, and the integral $\int d^{d-1}\vec{k}$ evidently vanishes. As a consequence, the whole expression is equal to zero.
	
	A straightforward calculation of the graph (b), within the dimensional regularization framework, yields:
	\bea
	I_b&=&-\frac{\xi}{2}\tilde{B}^{\mu}\tilde{B}^{\nu}\mu^{4-d}\int\frac{d^dk}{(2\pi)^d}\frac{k_{\mu}k_{\nu}}{k^2-m^2}+\xi\frac{m^2}{4}\tilde{B}^{\mu}\tilde{B}_{\mu}
	\mu^{4-d}\int\frac{d^dk}{(2\pi)^d}\frac{1}{k^2-m^2}+O(\xi^2)=\nonumber\\
	&=&\Gamma(1-\frac{d}{2})\frac{m^4\xi}{128\pi^2}\tilde{B}^{\mu}\tilde{B}_{\mu}+O(\xi^2),
	\eea
	which represents the mass renormalization for the vector field. For the graph (c) we have
	\bea
	I_c&=&\frac{\xi}{2}<\tilde{B}^{\mu}\tilde{B}^{\nu}>\partial_{\mu}\Phi\partial_{\nu}\Phi+\frac{m^2\xi}{4}<\tilde{B}_{\mu}\tilde{B}^{\mu}>\Phi\Phi=\nonumber\\
	&=&\frac{\xi}{2}(\partial^{\alpha}\Phi\partial^{\beta}\Phi+\frac{m^2}{2}\eta^{\alpha\beta}\Phi^2)\int\frac{d^4k}{(2\pi)^4}\frac{1}{k^2}\Big[\eta_{\alpha\beta}-
	\frac{1}{2\Lambda(\beta\cdot k)^2}\Big([-k^2-2\Lambda\beta^2]k_{\alpha}k_{\beta}\nonumber+\\
	\nonumber&+&2\Lambda(\beta\cdot k)(\beta_{\alpha}k_{\beta}+\beta_{\beta}k_{\alpha})
	\Big)\Big]=\nonumber\\
	&=&\frac{\xi}{4\Lambda}(\partial^{\alpha}\Phi\partial^{\beta}\Phi+\frac{m^2}{2}\eta^{\alpha\beta}\Phi^2)
	\int\frac{d^4k}{(2\pi)^4}
	\frac{k_{\alpha}k_{\beta}}{(\beta\cdot k)^2},
	\eea
	where we used the above integrals. Then, we use the identity
	$$
	\int\frac{d^4k}{(2\pi)^4}
	\frac{k_{\alpha}k_{\beta}}{(\beta\cdot k)^2}=-\frac{\partial^2}{\partial\beta^{\alpha}\partial\beta^{\beta}}
	\int\frac{d^4k}{(2\pi)^4}\ln(k\cdot\beta)=0,
	$$
	since we can choose e.g. $\beta^{\mu}=(\beta,0,0,0)$, and it is clear that $\int\frac{d^4k}{(2\pi)^4}\ln(k_0\beta)=0$. We note that Feynman graphs involving triple scalar-vector vertices yield contributions of the order $\xi^2$ and are irrelevant within the order of approximation of our analysis. 	
	

	In the case of a nontrivial vacuum, there will be a new contribution to the two-point function of the vector field generated by the Feynman diagram with two triple vertices each of which is proportional to $\Lambda$ (see Fig. 2).
	\begin{figure}[h]
		\centering
		\includegraphics[scale=1]{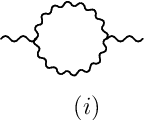}
		\caption{\label{Fig3} The new contribution to the two-point function.}
	\end{figure}
	This contribution was already studied in \cite{PJ2}, and given that it is  proportional to the integrals (\ref{kk},\ref{kkk})  (cf. \cite{Leib}), it was shown to be equal to zero. Therefore, the diagram (i) does not contribute to the effective potential. 
	
	\section{Effective potential}\label{potential}
	
	Let us now consider the effective potential in our theory focusing first on the vector field. Following general statements of the concept of the effective potential \cite{CW}, it can be calculated as the effective Lagrangian at slowly varying external fields. We start with the action (\ref{BumbLagVEV}). If we assume the scalar field $\Phi$ to be purely internal (just as the spinor in \cite{PJ3}), and the vector $B_{\mu}$ to be purely external, disregarding its derivatives, we can write down the effective action of the $B_{\mu}$ through the following functional trace:
	\bea
	\Gamma^{(1)}&=&\frac{i}{2}{\rm tr}\ln(\Box+m^2 -\xi[(B\cdot\partial)^2+\frac{m^2}{2}B^2]).
	\eea
	After Fourier transform and Wick rotation (both for momenta and fields), we write down, in $d$ dimensions,
	\bea
	\Gamma^{(1)}&=&-\frac{1}{2}\int\frac{d^dk_E}{(2\pi)^d}\ln(k^2_E+m^2(1+\frac{\xi B^2_E}{2}) +\xi(B_E\cdot k_E)^2).
	\eea
	Let us define also $\tilde{m}^2=m^2(1+\frac{\xi B^2_E}{2})$.
	We calculate this integral by two methods. 
	
	In the first one (cf. \cite{free}), we choose the background vector to be directed along one of coordinate axes, e.g. $B_{\mu}=(B,0,0,0)$:
	\bea
	\Gamma^{(1)}&=&-\frac{1}{2}\int\frac{d^dk_E}{(2\pi)^d}\ln(k^2_E+\tilde{m}^2 +\xi b^2_0k^2_{0E}).
	\eea
	We rescale $k_{0E}\sqrt{1+\xi B^2}=\tilde{k}_{0E}$, thus $d^dk=\frac{d^dk}{\sqrt{1+\xi B^2}}$, and $\tilde{k}_{0E}^2+\vec{k}^2=\tilde{k}^2$. So,
	\bea
	\Gamma^{(1)}&=&-\frac{1}{2}\frac{1}{\sqrt{1+\xi B^2}}\int\frac{d^d\tilde{k}_E}{(2\pi)^d}\ln(\tilde{k}^2_E+\tilde{m}^2).
	\eea
	The integral is straightforward. Introducing $\epsilon=2-\frac{d}{2}$ and using the relation $\Gamma(x+1)=x\Gamma(x)$, we can write (omitting the index $E$ for compactness)
	\bea
	\label{res1}
	\Gamma^{(1)}&=&-\frac{1}{32\pi^2}\frac{1}{\sqrt{1+\xi B^2}}\frac{\Gamma(2-\frac{d}{2})}{d(1-\frac{d}{2})}\tilde{m}^4\left(\frac{\tilde{m}^2}{4\pi\mu^2}\right)^{d/2-2}=\nonumber\\
	&=&\frac{m^4}{64\pi^2}\frac{(1+\xi B^2/2)^2}{\sqrt{1+\xi B^2}}
	\Big[\frac{2}{\epsilon}+\gamma+\frac{3}{2}-\ln(\frac{m^2}{\mu^2}(1+\frac{\xi B^2}{2}))\Big]+O(\epsilon).
	\eea
	
	In the second one (see e.g. \cite{average}), we use `averaging" by the rule $k_{\mu}k_{\nu}=\frac{1}{4}\delta_{\mu\nu}k^2_E$, so,
	\bea
	\Gamma^{(1)}&=&-\frac{1}{2}\int\frac{d^dk_E}{(2\pi)^d}\ln(k^2_E+\tilde{m}^2+\frac{\xi}{4}B^2_Ek^2_E)=
	-\frac{1}{2}\int\frac{d^dk_E}{(2\pi)^d}\ln(k^2_E(1+\frac{\xi B^2_E}{4})+\tilde{m}^2),
	\eea
	afterwards, we change $k^2_E(1+\frac{\xi B^2_E}{4})=\tilde{k}^2$, so, $d^4k=\frac{d^4\tilde{k}}{(1+\frac{\xi B^2_E}{4})^2}$, hence (omitting the index $E$ again)
	\bea
	\label{res2}
	\Gamma^{(1)}&=&-\frac{1}{2}\frac{1}{(1+\frac{\xi B^2_E}{4})^2}\int\frac{d^d\tilde{k}_E}{(2\pi)^d}\ln(\tilde{k}^2_E+\tilde{m}^2)=\nonumber\\
	&=&\frac{1}{32\pi^2}\frac{1}{(1+\xi B^2/4)^2}\frac{\Gamma(2-\frac{d}{2})}{d(1-\frac{d}{2})}\tilde{m}^4\left(\frac{\tilde{m}^2}{4\pi\mu^2}\right)^{d/2-2}=\nonumber\\
	&=&\frac{m^4}{64\pi^2}\frac{(1+\xi B^2/2)^2}{(1+\xi B^2/4)^2}
	\Big[\frac{2}{\epsilon}+\gamma+\frac{3}{2}-\ln(\frac{m^2}{\mu^2}(1+\frac{\xi B^2}{2}))\Big]+O(\epsilon).
	\eea
	We note that the only difference between the results (\ref{res1},\ref{res2}) is in their denominators: 
	$(1+\xi B^2)^{1/2}$ and $(1+\xi B^2/4)^2$ respectively. However, since $(1+\xi B^2)^{1/2}\simeq 1+\frac{\xi}{2}B^2+O(\xi^2)$, and 
	$(1+\xi B^2/4)^2\simeq 1+\frac{\xi}{2}B^2+O(\xi^2)$, we conclude that these results are equivalent at first order in $\xi$ (which is the only interesting one for us). 
	
	As a next step, one can try to obtain the effective potential of the $\Phi$ field, which becomes a really interesting task namely for $\beta_{\mu}\neq 0$. In this case, the corresponding expression is given by
	\bea
	\Gamma^{(1)}[\Phi]&=&\frac{i}{2}\int\frac{d^4k}{(2\pi)^4}\ln\det\Big( (-k^2\eta_{\mu\nu}+k_{\mu}k_{\nu})(1+\frac{\xi\beta^2}{2})+\xi(k^2\beta_{\mu}\beta_{\nu}+(\beta\cdot k)^2\eta_{\mu\nu})-2\Lambda\beta_{\mu}\beta_{\nu} +\nonumber\\&+&
	\frac{m^2\xi}{2}\Phi^2\eta_{\mu\nu}
	\Big).
	\eea
	This expression is very cumbersome. Actually, for the simplest case of $\beta_{\mu}=\beta\eta_{\mu 0}$ (i.e. $\beta_{\mu}$ is directed along the time axis) we can calculate the determinant easily arriving at
	\bea
	\Gamma^{(1)}[\Phi]&=&\frac{i}{2}\int\frac{d^4k}{(2\pi)^4}\ln\Big( -[-k^2+\xi\beta^2k^2_0+\frac{\xi m^2\Phi^2}{2}]^4-\nonumber\\
	&-&(
	(1+\frac{\xi\beta^2}{2})k^2-2\Lambda\beta^2+\xi\beta^2  k^2)[-k^2+\xi\beta^2k^2_0+\frac{\xi m^2\Phi^2}{2}]^3+\nonumber\\&+&
	(1+\frac{\xi\beta^2}{2})(-2\Lambda\beta^2+\xi \beta^2 k^2)\vec{k}^2[-k^2+\xi\beta^2k^2_0+\frac{\xi m^2\Phi^2}{2}]^2
	\Big),
	\eea
	which, up to the irrelevant additive constant, can be written as
	\bea
	\Gamma^{(1)}[\Phi]&=&i\int\frac{d^4k}{(2\pi)^4}\ln[-k^2+\xi\beta^2k^2_0+\frac{\xi m^2\Phi^2}{2}]+\nonumber\\&+&
	\frac{i}{2}\int\frac{d^4k}{(2\pi)^4}\ln\Big( [-k^2+\xi\beta^2k^2_0+\frac{\xi m^2\Phi^2}{2}]^2+\\
	&+&(
	(1+\frac{\xi\beta^2}{2})k^2-2\Lambda\beta^2+\xi\beta^2  k^2)[-k^2+\xi\beta^2k^2_0+\frac{\xi m^2\Phi^2}{2}]-\nonumber\\&-&
	(1+\frac{\xi\beta^2}{2})(-2\Lambda\beta^2+\xi \beta^2 k^2)\vec{k}^2
	\Big)=\Gamma_1+\Gamma_2, \nonumber
	\eea
	where $\Gamma_1$ and $\Gamma_2$ stand for the two integrals involved in this expresion. While the first integral can be calculated easily, the second one apparently cannot be simplified essentially and will be processed in an approximate manner. To proceed, we employ the standard Euclidean integral:
	\bea
	I(M^2)=\mu^{-\epsilon}\int\frac{d^{4+\epsilon}k_E}{(2\pi)^{4+\epsilon}}\ln(k^2_E+M^2)=\frac{M^4}{16\pi^2}(\frac{2}{\epsilon}+\ln\frac{M^2}{\mu^2})+O(\epsilon).
	\eea
	Similarly, in the Euclidean space, doing some redefinitions, we can write
	\bea
	I(M^2)=\mu^{-\epsilon}\int\frac{d^{4+\epsilon}k_E}{(2\pi)^{4+\epsilon}}\ln(ak^2_{0E}+b\vec{k}^2+M^2)=\frac{M^4}{16\pi^2(ab^3)^{1/2}}(\frac{2}{\epsilon}+\ln\frac{M^2}{\mu^2})+O(\epsilon).
	\eea
	Hence, after rescaling $k_0$, we have
	\bea
	\Gamma_1&=&i\int\frac{d^4k}{(2\pi)^4}\ln[-k^2+\xi\beta^2k^2_0+\frac{\xi m^2\Phi^2}{2}]=\nonumber\\ &=&
	-\frac{1}{\sqrt{1-\xi\beta^2}}\frac{1}{16\pi^2}\left(\frac{\xi m^2\Phi^2}{2}\right)^2\left[\frac{2}{\epsilon}+\ln\left(\frac{\xi m^2\Phi^2}{2\mu^2}\right)\right].
	\eea
	Then,
	\bea
	\Gamma_2&=&\frac{i}{2}\int\frac{d^4k}{(2\pi)^4}\ln\Big( [-k^2+\xi\beta^2k^2_0+\frac{\xi m^2\Phi^2}{2}]^2+\nonumber\\
	&+&(
	(1+\frac{\xi\beta^2}{2})k^2-2\Lambda\beta^2+\xi\beta^2  k^2)[-k^2+\xi\beta^2k^2_0+\frac{\xi m^2\Phi^2}{2}]-\nonumber\\&-&
	(1+\frac{\xi\beta^2}{2})(-2\Lambda\beta^2+\xi \beta^2 k^2)\vec{k}^2
	\Big)=\nonumber\\
	&=&\frac{i}{2}\int\frac{d^4k}{(2\pi)^4}\ln\Big[
	k^4_{0E}(-\frac{5}{2}\xi\beta^2+\frac{5}{2}\xi^2\beta^4)+k^2_{0E}\vec{k}^2(-\frac{3}{2}\xi\beta^2+2\xi^2\beta^4)+\\&+& \vec{k}^4(-\frac{1}{2}\xi\beta^2+\frac{1}{2}\xi^2\beta^4)-
	\beta^2k^2_{0E}(\frac{5}{4}\xi^2 m^2\Phi^2+2\Lambda(1-\xi\beta^2))-\beta^2\vec{k}^2(\frac{\xi^2 m^2\Phi^2}{4}+2\Lambda)
	\Big].\nonumber
	\eea
	This integral is very complicated. However, if we assume that $\xi\beta^2\ll 1$ (which is natural given that both the nonminimal coupling constant and the LV vector are small), and subtract some constants (namely, $\ln(\xi\beta^2)$ and $(-1)$ multiplier), we have
	\bea
	\Gamma_2&=&\frac{i}{2}\int\frac{d^4k}{(2\pi)^4}\ln\Big[
	k^4_{0E}(\frac{5}{2}\xi\beta^2)+k^2_{0E}\vec{k}^2(\frac{3}{2}\xi\beta^2)+\vec{k}^4(\frac{1}{2}\xi\beta^2)+\nonumber\\&+&
	\xi \beta^2k^2_{0E}(\frac{5}{4}\xi m^2\Phi^2+2\frac{\Lambda}{\xi})+\xi\beta^2\vec{k}^2(\frac{\xi m^2\Phi^2}{4}+2\frac{\Lambda}{\xi})
	\Big]=\nonumber\\
	&=&\frac{i}{2}\int\frac{d^4k}{(2\pi)^4}\ln\Big[
	k^4_{0E}(\frac{5}{2})+k^2_{0E}\vec{k}^2(\frac{3}{2})+\vec{k}^4(\frac{1}{2})+\nonumber\\&+&
	k^2_{0E}(\frac{5}{4}\xi m^2\Phi^2+2\frac{\Lambda}{\xi})+\vec{k}^2(\frac{\xi m^2\Phi^2}{4}+2\frac{\Lambda}{\xi})
	\Big]
	\eea
	This integral (and consequently the result) is still rather cumbersome. For this reason, and in order to describe its general dependence on parameters of the theory, we suppress all numerical coefficients of order 1 replacing them just by 1 and, after subtracting the divergence, we arrive at
	\bea
	\Gamma_2&=&\frac{i}{2}\int\frac{d^4k}{(2\pi)^4}\ln\Big[k^4_E+
	k^2_E(\xi m^2\Phi^2+\frac{\Lambda}{\xi})
	\Big]=\nonumber\\
	&=&-\frac{1}{32\pi^2}\frac{\Lambda^2}{\xi^2}(1+\frac{\xi^2 m^2\Phi^2}{\Lambda})^2[\ln\frac{\Lambda}{\xi\mu^2}+\ln(1+\frac{\xi^2 m^2\Phi^2}{\Lambda})].
	\eea
	We note that, unlike $\Gamma_1$, the smallness of this result is characterized by the relation $\frac{\xi^2m^2\Phi^2}{\Lambda}$: if $\Lambda$ is not extremely small, we have $\frac{\xi^2m^2\Phi^2}{\Lambda}\ll 1$ in physically reasonable situations.
	
	A comment is in order here. The above result can be naively treated to be ill-defined at $\xi\to 0$. However, if we expand it in powers of $\xi$, we will first see that the $\xi^{-2}$ term actually  does not depend on background fields and gives no physical contribution to the effective potential, and second, that the term of zero order in $\xi$, which looks like $-\frac{1}{16\pi^2}(\Lambda^2\ln\frac{\Lambda}{\xi\mu^2}+\frac{1}{2}\Lambda) m^2\Phi^2$, vanishes for a certain choice of the normalization parameter $\mu$, being therefore an artifact of the calculation scheme. Thus, physical contributions indeed begin at order $\xi^2$, as noted earlier. We note nevertheless that the singularity of our result at $\xi\to 0$ illustrates the fact that in this limit our vector-scalar vertex vanishes making the calculation meaningless.
	
	Thus, at $\xi\beta^2\ll 1$ (so, the effective potential does not depend on the LV vector but depends on a small parameter describing nonminimal bumblebee-gravity coupling), after subtracting divergences, we have
	\bea
	\Gamma^{(1)}&=&\Gamma_1+\Gamma_2\simeq
	-\frac{1}{16\pi^2}\left(\frac{\xi m^2\Phi^2}{2}\right)^2\ln\left(\frac{\xi m^2\Phi^2}{2\mu^2}\right)-\nonumber\\
	&-&\frac{1}{32\pi^2}\frac{\Lambda^2}{\xi^2}(1+\frac{\xi^2 m^2\Phi^2}{\Lambda})^2[\ln\frac{\Lambda}{\xi\mu^2}+\ln(1+\frac{\xi^2 m^2\Phi^2}{\Lambda})].
	\eea
	In principle, the physical impacts of this potential can be studied in detail. We note again that the lower nontrivial contribution to the effective potential is indeed proportional to $\xi^2$, as we claimed in the previous section.

	\section{Summary and conclusions}\label{con}
	
	In this work we have studied the perturbative impacts of coupling the metric-affine bumblebee gravity to scalar matter, a model introduced originally in \cite{PJ1}. For this theory, we calculate the two-point functions and effective potentials of vector and scalar fields. The purely vector sector of the theory had been treated earlier in \cite{PJ1}, where it was shown that all contributions of zero order in $\xi$ vanish, so that the effective action of the $B_{\mu}$ field, with its couplings to other fields switched off, begins with the first order in $\xi$. Its study requires rather involved calculations, which we plan to consider elsewhere. Nevertheless, by dimensional reasons it is natural to expect divergent contributions like $\xi\beta^2 B^4$ or $\xi (\beta\cdot B)^2B^2$, and the corresponding finite logarithmic-like results. At the same time, the contribution of the first order in $\xi$ to the effective action of the background scalar field was shown to vanish, so that this effective action will begin with $\xi^2$.
	
	The main focus of the paper was the calculation of the effective potential, which was considered in two situations, namely, i) the effective potential of the vector field when the scalar is integrated out, and ii) when the vector field is integrated out and the effective potential of the scalar field is found. In the first case, we obtain the result with the use of two schemes, with the first one based on fixing the direction of the LV vector along one of the coordinate axes, and the other based on some averaging over space-time directions  (as a by-product, we showed their equivalence in the first order expansion in the small parameter $\xi$), and the result can be presented as an infinite series in the scalar $\xi B^2$, where $B_{\mu}$ is the background field. We further applied these results to studies of the dynamical Lorentz symmetry breaking in our theory.  In the second case, we calculated the effective potential of the scalar field, which, under certain normalization prescription, turns out to begin with the $\xi^2$ order. We note that in principle this calculation is easily generalized for the case of $N$ scalar fields which allows us to study dynamical breaking of the corresponding isotopic symmetry.
	
	Also, it is instructive to do some numerical estimations for impacts of the Lorentz symmetry breaking in our theory. The effects of Lorentz-symmetry breaking are characterized by the dimensionless parameter $X=\xi \beta^2$. As $X$ increases, departures from Lorentz-invariant theories become more evident, making clear the significance of accurately estimating this parameter through experimental data. The $\beta_{\mu}$ is the norm-fixed VEV vector whose norm is $\beta$, i.e., it can be treated as a scale for some LV vector parameter. In \cite{datatables}, the upper scale for any constant vector is found to be, from various physical situations within the LV context, no more than $10^{-21}$ GeV. Estimating solely $\xi$ is challenging due to the absence of experimental data specifically aimed at this purpose, where $\xi$ essentially represents the inverse squared energy scale of nonminimal coupling. However, let us assume that this energy scale is much less than known Planck and string scales, i.e. that this is the characteristic scale corresponding to $R^2$ terms within the gravity Lagrangian. It has been estimated that the constant $M^2$ corresponding to the $R^2$ gravity, with ${\cal L}=R+\frac{R^2}{6M^2}$, is about $10^{-11}$ GeV \cite{Cemb,Berry}. If we assume $\xi\simeq M^{-2}$, we obtain $|\xi b^2|\leq 10^{-20}$. While this is a very tiny number, probably it can impact in certain physical situations like fine-tuning effects.
	
	A natural continuation of this study, besides calculating higher-order contributions to the effective potential and investigating the dynamical symmetry breaking, may consider the introduction of a nontrivial, non-flat gravitational background. We plan to perform these studies in forthcoming papers.
	
	\acknowledgments
	
	This work was supported by Conselho Nacional de Desenvolvimento Cient\'{\i}fico e Tecnol\'{o}gico (CNPq). PJP would like to thank the Brazilian agency CNPq for financial support (PQ--2 grant, process 
	No. 307628/2022-1).  The work by A. Yu. P. has been supported by the CNPq project No. 303777/2023-0. This work is also supported by the Spanish Agencia Estatal de  Investigaci\'on (grant PID2020-116567GB-C21 funded by MCIN/AEI/10.13039/501100011033 and ERDF A way of making Europe), and by the project PROMETEO/2020/079 (Generalitat Valenciana).
	


\end{document}